\documentclass[preprint,
superscriptaddress,
amsmath,amssymb,
aps,
prl,
floatfix,
]{revtex4-2}

\usepackage{natbib,hyperref}
\usepackage{graphicx}
\usepackage{dcolumn}
\usepackage{bm}
\usepackage[colorinlistoftodos]{todonotes}
\usepackage{xcolor}
\usepackage{hyperref}  
\usepackage[utf8]{inputenc}
\usepackage{amsmath}
\setcitestyle{numbers,square}
\usepackage{setspace}
\usepackage{indentfirst}

\DeclareGraphicsExtensions{.png,.pdf} 
\begin{document}

\title{Extremely thin perfect absorber by generalized multipole bianisotropic effect}


\author{Hao Ma}
\affiliation{Research Institute of Superconductor Electronics (RISE) \& Key Laboratory of Optoelectronic Devices and Systems with Extreme Performances of MOE, School of Electronic Science and Engineering, Nanjing University, Nanjing 210023, China}
\affiliation{Purple Mountain Laboratories, Nanjing 211111, China}
\author{Andrey B. Evlyukhin}
\affiliation{Institute of Quantum Optics, Leibniz Universität Hannover, Welfengarten 1, 30167  Hannover, Germany}

\author{Andrey E. Miroshnichenko}
\affiliation{School of Engineering and Information Technology, University of New South Wales at Canberra, Northcott Drive, Canberra ACT 2610, Australia}
\author{Fengjie Zhu}
\affiliation{Research Institute of Superconductor Electronics (RISE) \& Key Laboratory of Optoelectronic Devices and Systems with Extreme Performances of MOE, School of Electronic Science and Engineering, Nanjing University, Nanjing 210023, China}
\author{Siyu Duan}
\affiliation{Research Institute of Superconductor Electronics (RISE) \& Key Laboratory of Optoelectronic Devices and Systems with Extreme Performances of MOE, School of Electronic Science and Engineering, Nanjing University, Nanjing 210023, China}
\author{Jingbo Wu}
\affiliation{Research Institute of Superconductor Electronics (RISE) \& Key Laboratory of Optoelectronic Devices and Systems with Extreme Performances of MOE, School of Electronic Science and Engineering, Nanjing University, Nanjing 210023, China}
\affiliation{Purple Mountain Laboratories, Nanjing 211111, China}
\author{Caihong Zhang}
\affiliation{Research Institute of Superconductor Electronics (RISE) \& Key Laboratory of Optoelectronic Devices and Systems with Extreme Performances of MOE, School of Electronic Science and Engineering, Nanjing University, Nanjing 210023, China}
\affiliation{Purple Mountain Laboratories, Nanjing 211111, China}
\author{Jian Chen}
\affiliation{Research Institute of Superconductor Electronics (RISE) \& Key Laboratory of Optoelectronic Devices and Systems with Extreme Performances of MOE, School of Electronic Science and Engineering, Nanjing University, Nanjing 210023, China}
\affiliation{Purple Mountain Laboratories, Nanjing 211111, China}
\author{Biao-Bing Jin}
\email{bbj@nju.edu.cn}
\affiliation{Research Institute of Superconductor Electronics (RISE) \& Key Laboratory of Optoelectronic Devices and Systems with Extreme Performances of MOE, School of Electronic Science and Engineering, Nanjing University, Nanjing 210023, China}
\affiliation{Purple Mountain Laboratories, Nanjing 211111, China}
\author{Willie J. Padilla}
\email{willie.padilla@duke.edu}
\affiliation{Duke University, Department of Electrical and Computer Engineering, Durham, NC 2770 USA}

\author{Kebin Fan}
\email{kebin.fan@nju.edu.cn}
\affiliation{Research Institute of Superconductor Electronics (RISE) \& Key Laboratory of Optoelectronic Devices and Systems with Extreme Performances of MOE, School of Electronic Science and Engineering, Nanjing University, Nanjing 210023, China}
\affiliation{Purple Mountain Laboratories, Nanjing 211111, China}



\begin{abstract}

Symmetry breaking plays a crucial role in understanding the fundamental physics underlying numerous physical phenomena, including the electromagnetic response in resonators, giving rise to intriguing effects such as directional light scattering, supercavity lasing, and topologically protected states. In this work, we demonstrate that adding a small fraction of lossy metal (as low as $1\times10^{-6}$ in volume), to a lossless dielectric resonator breaks inversion symmetry thereby lifting its degeneracy, leading to a strong bianisotropic response. In the case of the metasurface composed of such resonators, this effect leads to unidirectional perfect absorption while maintaining nearly perfect reflection from the opposite direction. We have developed more general Onsager-Casimir relations for the polarizabilities of particle arrays, taking into account the contributions of quadrupoles, which shows that bianisotropy is not solely due to dipoles, but also involves high-order multipoles. Our experimental validation demonstrates an extremely thin terahertz-perfect absorber with a wavelength-to-thickness ratio of up to 25,000, where the material thickness is only 2\% of the theoretical minimum thickness dictated by the fundamental limit. Our findings have significant implications for a variety of applications, including energy harvesting, thermal management, single-photon detection, and low-power directional emission. 

\end{abstract}

\maketitle


\section{Introduction}
Metamaterials have brought about a paradigm shift in the approach to the manipulation of electromagnetic waves by allowing for interaction not only with the electric component of light but also with the magnetic component, which was previously challenging at high frequencies. The subwavelength nature of metamaterial unit-cells, i.e., meta-atoms, allows their scattering to be well described by the response of electric dipoles (EDs) and magnetic dipoles (MDs). The geometry of the meta-atoms enables the independent control of the ED ($\bm{p}$) and MD ($\bm{m}$) moments, showcasing its versatility in shaping the electromagnetic responses, including transparency \cite{Pfeiffer2013prl, Decker2015adom, Yu2015lpr, Abujetas2020lpr}, perfect absorption\cite{Liu2017oe, Fan2018am, Fan2021oe}, directional scattering, \cite{PaniaguaDomnguez2016nc, yang2017prb}, anapole effects \cite{Miroshnichenko2015nc} and exceptional points \cite{Moritake2023acsphoton}.

However, the constitutive equations indicate that the electric polarization and magnetization may not be inherently independent of each other. A possible cross-coupling between polarizations is termed bianisotropic polarizability (BAP) and enables coupling of electric to magnetic polarization ($\alpha_{em}$) and magnetic to electric polarization ($\alpha_{me}$), thereby providing an additional degree of freedom for manipulating electromagnetic waves. The bianisotropic effect can be induced by breaking the symmetry of metasurfaces (MSs), therefore leading to fascinating phenomena, such as asymmetric absorption, asymmetric reflection, enhanced harmonics generation, asymmetric parametric generation, and trapped modes with high-Q resonance \cite{Yazdi2015tap, Radi2015prappl, Wang2018prl, asadchy2015prl, Guo2016acsphoton, Cui2018acsphoton, Mobini2021acsphoton, Kruk2022np, Campione2016acsphoton, Evlyukhin2020prb}. Further, the reciprocity theorem requires $\overline{\overline{\alpha}}_{me} = -\overline{\overline{\alpha}}_{em}^T$, which is called the Onsager-Casimir condition, where $T$ denotes the transpose operation. However, a significant geometric change is often required to achieve a large asymmetric response \cite{Yazdi2015tap, Evlyukhin2020prb, Pfeiffer2014prl, Alaee2015, Alaee2015prb, Asadchy2018nanopho, alexei2022prb}. For instance, to obtain an asymmetric absorption, in one study the volume difference between the top and bottom patterns had to exceed 40\% \cite{Yazdi2015tap, Alaee2015prb}. An added complexity is that for structures which lack inversion symmetry (IS), fabrication can be challenging.

Although the behavior of MSs can often be well described by dipoles,  meta-atoms may not always be deeply subwavelength in the short wavelength range. Therefore, a full description of the resonator scattering needs to go beyond the simple dipole approximation, i.e., high-order multipoles should be included. Controlling the interaction of dipoles and high-order multipoles has enabled a new route to shape electromagnetic waves, such as the generalized Kerker effect, hybrid anapoles for transparency, and bound state in the continuum \cite{Fan2021oe, Babicheva2017prb, Liu2018oe, He2018prb, Shamkhi2019prm, Shamkhi2019prl, Sadrieva2019prb}. Further, the coupling among high-order multipoles could provide more physical effects such as  trapped multipole modes \cite{prokhorov2022resonant}, exceptional points \cite{valero2022magnetoelectric}, and the lattice anapole effect \cite{terekhov2019prb}.

In this letter, we present an extremely thin asymmetric perfect absorber which exhibits an extreme bianisotropic response far exceeding simple dipole coupling. By breaking the IS with an extremely subwavelength lossy disk on top of a lossless cylinder as illustrated in \textbf{Figure \ref{fig1}a}, the MS also exhibits nearly perfect reflection under backward illumination. The developed theory reveals that the bianisotropy is not only caused by dipole cross-coupling, but also enhanced by quadrupoles. We experimentally demonstrate an asymmetric terahertz perfect absorber by patterning lossy Niobium nitride (NbN) disks on dielectric cylinders as shown in the inset of Figure \ref{fig1}b. The patterned disk absorbs 99.999\% of the power, and achieves a remarkable wavelength-to-thickness ratio of 25,000, thereby surpassing the theoretically established limits of absorber thickness. Further, the volume filling fraction of the demonstrated absorber is two orders of magnitude smaller than other absorbing thin films in literature, as shown in Figure \ref{fig1}b. This breakthrough not only illuminates new pathways for exploring light-matter interactions but also promises a new generation of high-performance metasurface devices.

\section{Results \& Discussion}

\begin{figure}[tb]
	\centering
	\includegraphics[width = 16cm]{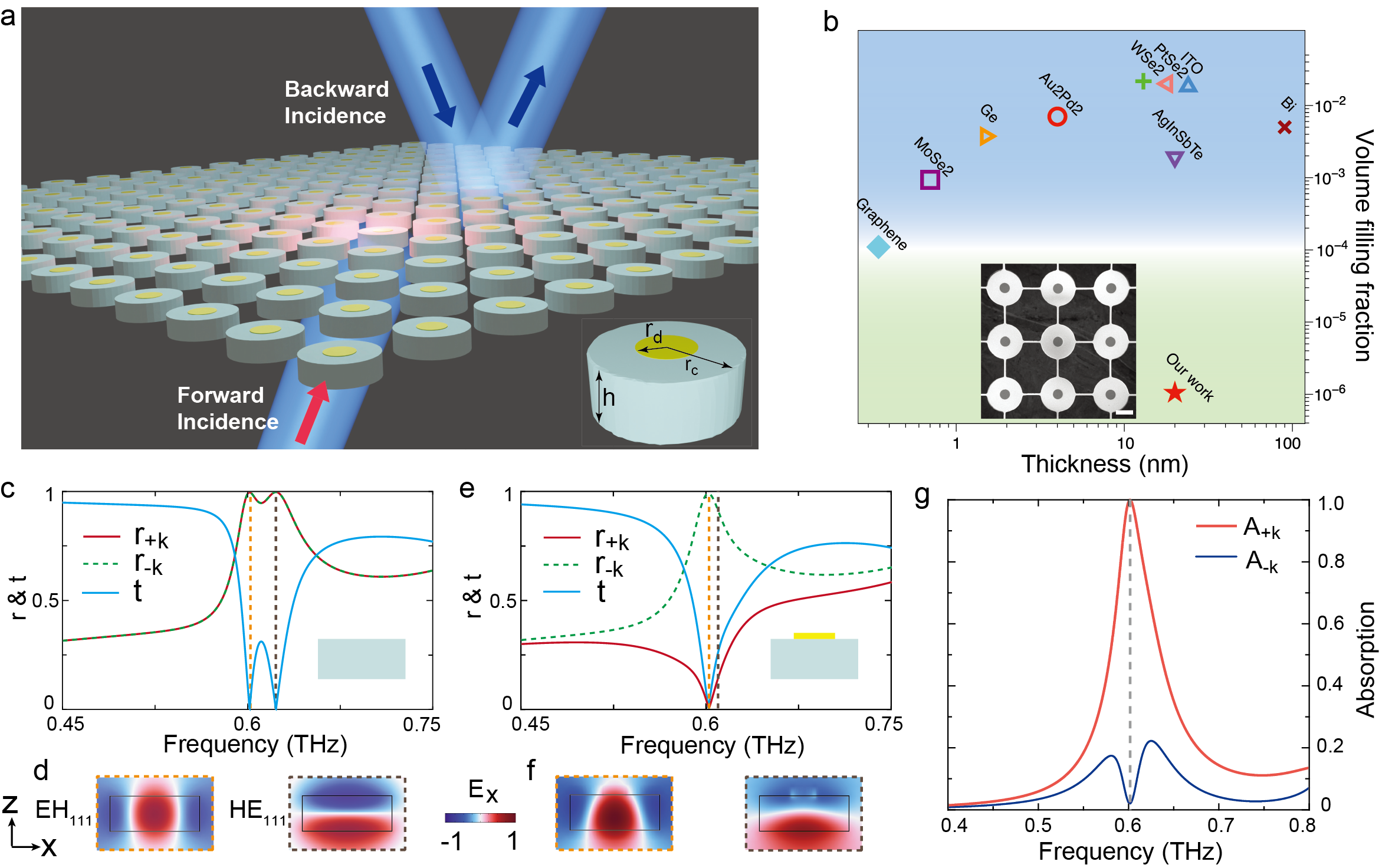}
	\caption{Concept illustration and comparison of demonstrated absorber performance with other literature. (a) Illustration of the extremely thin absorber consisting of an array of lossless dielectric cylinders with metallic disks patterned on the top. The structure shows perfect absorption under forward illumination (red arrow) and nearly perfect reflection under backward illumination. (b) Comparison of the volume filling fraction of our absorbing disk and other thin-film absorbers, including graphene \cite{Nematpouranm2021}, MoS$_2$ \cite{Horng2020prappl},  Ge \cite{songadam2014}, Au$_2$Pd$_2$ \cite{Xiaoanm2021}, WeSe$_2$ \cite{Jariwalanl2016}, PtSe$_2$ \cite{Alfieriadom2022}, AgInSbTe \cite{heleraom2021}, ITO \cite{lukprb2014}, Bi \cite{Toudertoe2018}. The inset shows a scanning electron microscopic image of the fabricated free-standing asymmetric metasurface, where the scale bar is 100 $\mu$m. c)-f) Comparison between metasurfaces with (c,d) and without (e,f) IS. (c) \& (e) Simulated transmission and reflections under the forward ($+k$) and backward ($+k$) illuminations. The dimensions of the cylinders are $r_c=103$ $\mu$m, $h=80$ $\mu$m and $p=350$ $\mu$m. The asymmetric MS in (e) is achieved by patterning 4.4-nm thick metallic disks with $r_d=28.5$ $\mu$m on top of the cylinders. The insets in (c) and (e) illustrate the symmetric and asymmetric metasurfaces.  (d) \& (f) shows the electric field ($E_x$) distributions at the two eigenmodes at the frequencies indicated by the dashed orange and gray dashed vertical lines, respectively, of (c) and (e). (g) Simulated absorption spectra under $+k$ (red) and $-k$ (black) illuminations with the $r_d =  28.5 \mu$m.}
	\label{fig1}
\end{figure}

To obtain asymmetric perfect absorption from an initially symmetric MS, different reflection under forward and backward incidence has to be achieved. Here, we break IS through the introduction of a lossy conductive disk on top of a cylindrical resonator, as depicted in Figure \ref{fig1}a. In Figure\ref{fig1}c and e, we compare the simulated scattering response of a metasurface with and without IS. For MSs consisting of lossless silicon cylinders with IS about its center point, the reflections with illumination from the top ($-k$) direction (dashed green curve) and bottom ($+k$) direction (solid red curve) are identical, as shown in Figure\ref{fig1}(b). The transmission remains the same due to reciprocity. In this case, two fundamental modes, namely the EH$_{111}$ and HE$_{111}$ modes \cite{Liu2017oe}, are excited with their electric field distributions shown in Figure \ref{fig1}d. The symmetric and anti-symmetric profiles of the modes indicate that there is no cross-coupling between the scattering responses with even and odd parity. However, for the broken symmetry case where a 4.4-nm thick lossy disk with a radius of 28.5 $\mu$m is added to one side (Figure \ref{fig1}e), the forward reflection $r_{+k}$ is significantly suppressed, while the backward reflection $r_{-k}$ remains close to 1, as shown in Figure\ref{fig1}e. Notably, despite the conductive disk's thickness being a mere 8.8$\times10^{-6}$ fraction of the operational wavelength of 500 $\mu$m, its influence on the bianisotropic response is profound. Here the disk is modeled as gold with a conductivity of $\sigma = 2\times 10^6$ S/m, which is over one order of magnitude lower than its bulk value, given its few-nanometer thickness \cite{Lee2019sr}. Because of the conductive layer, the electric field at the top surface of the cylinder is partially screened, leading to a significantly damped resonance at 0.61 THz, as shown in Figure \ref{fig1}f. In this case, the field of the first mode near 0.6 THz changes from an ellipse-like shape to a cone-like shape, and the zero electric field plane of the second mode near 0.61 THz shifts lower, as shown in Figure \ref{fig1}f. In comparison to the symmetric case, the asymmetric field distribution indicates that the scattering response consists of mixed even and odd parity states.

\begin{figure}[tp]
	\centering
	\includegraphics[width= 16cm]{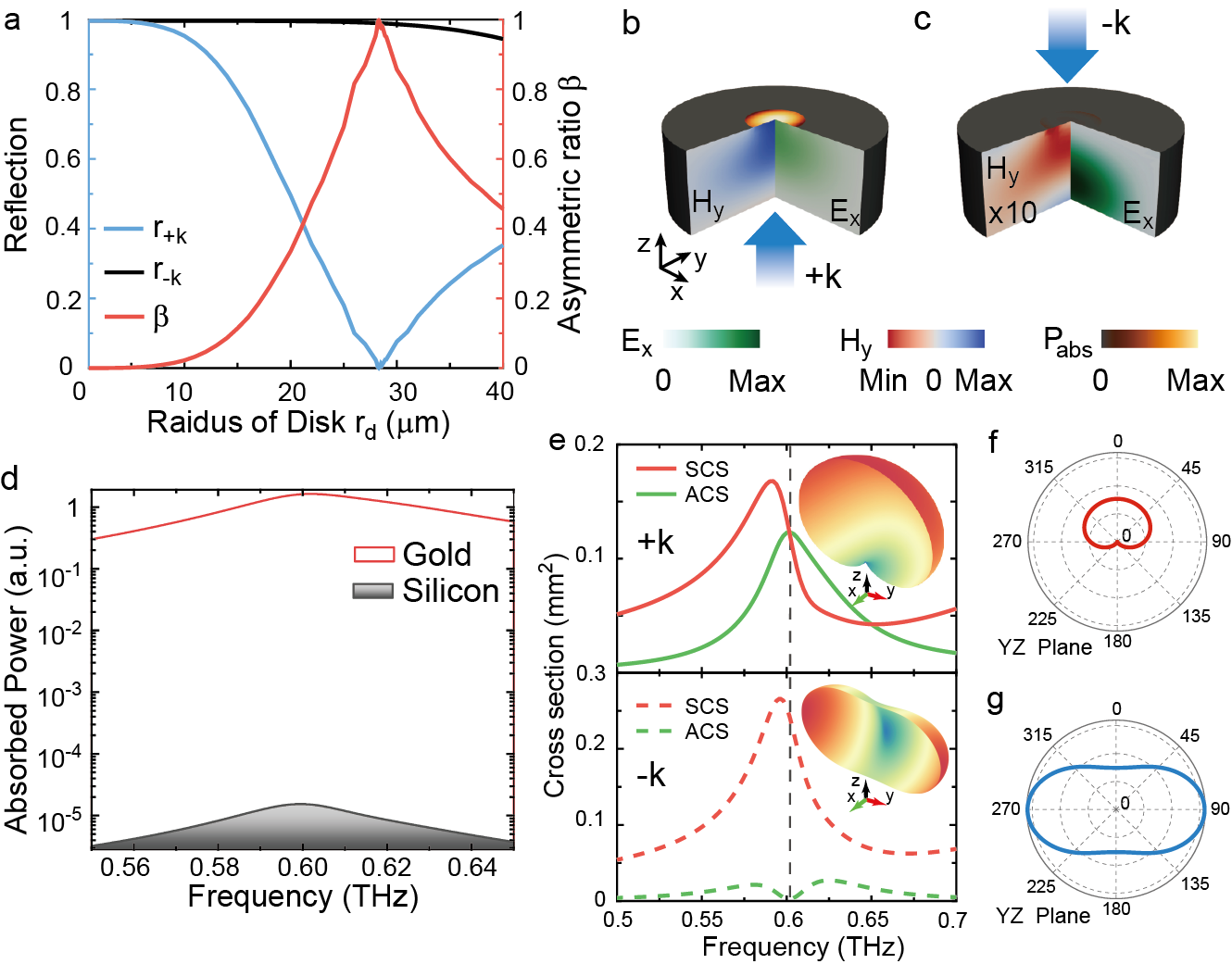}
	\caption{Asymmetric response of the symmetry broken metasurface. (a) The asymmetric reflections and the asymmetric ratio varying with the disk radius.(b) \& (c) The distribution of the electric field E$_x$ in the ZX plane, the magnetic field H$_y$ in the ZY plane, and the power loss density in the disk under the $+k$ (b) and $-k$ (c) illuminations at 0.6 THz. (d) Calculated absorbed power in the disk (red) and the cylinder (gray), respectively. (e) simulated total scattering cross sections (SCS) and total absorption cross sections (ACS) under the $+k$ (top panel) and $-k$ (bottom panel) illuminations for the structure with r$_d$ = 28.5 $\mu$m. The insets show the sliced radiation patterns along the YZ plane, for two opposite incidences, respectively.(f) \& (g) show the corresponding radiation patterns at the YZ plane (H plane) for the two cases, respectively. }
	\label{fig2}
\end{figure}

The geometry shown in Figure \ref{fig1}e permits the asymmetric metasurface to achieve perfect absorption ($A_{+k}$) under $+k$ illumination and nearly zero absorption under $-k$ incidence as shown in Figure \ref{fig1}g; a result attributable to the symmetry breaking introduced by the patterned conductive disk. The absorption is strongly influenced by both the structural geometry and the disk's conductivity -- details can be found in Section 4 of the Supporting Information. In \textbf{Figure \ref{fig2}a}, the reflection coefficients ($r_{+k}$, $r_{-k}$) are illustrated, along with their dependence on disk radius at 0.6 THz. To assess the variation between the reflection from both sides, we establish an asymmetric factor, denoted as $\beta=(|r_{-k} |-|r_{+k} |)/(|r_{+k} |+|r_{-k} |)$. With a smaller disk radius, both $r_{+k}$ and $r_{-k}$ approach unity, thereby resulting in a nearly null asymmetric factor. However, for $r_d$ exceeding 10 $\mu$m, the forward reflection $r_{+k}$ drops precipitously to zero while the backward reflection $r_{-k}$ remains over 0.94. Here, the asymmetric factor ascends to its maximum of 1 at $r_d$= 28.5 $\mu$m.

It is crucial to underscore that achieving a nearly perfect asymmetric factor with such a minimal volume filling fraction (VFF) -- here a value of 1.8$\times10^{-7}$ -- is unprecedented. In this context, VFF is expressed as $\rm{VFF} = t_{d}S_{d}/\lambda^3$, where $t_{d}$ denotes the thickness of the absorbing film, and $S_{d}$ indicates the film's surface area within one wavelength area ($\lambda^2$). Given the negligible change in transmission at 0.6 THz, the structure exhibits perfect absorption under $+k$ directional illumination and nearly null absorption ($A_{-k}$) under the $-k$ illumination at maximum $\beta$, as evidenced in Figure \ref{fig1}g. This asymmetric perfect absorption ties directly into the asymmetric field distribution within the structure. Under the $+k$ illumination, the excited electric and magnetic field at 0.6 THz predominantly localizes near and within the disk (see Figure \ref{fig2}b). As the disk is the sole lossy component in the structure, the incident power will largely be dissipated within it as $P_{abs}=\frac{1}{2}\int \sigma|E|^2 dV$, leading to significant energy confinement within the thin film.

Given that intrinsic silicon has an electron concentration of 10$^{11}$ /cm$^3$, we confirm that the conductive disk absorbs an astounding 99.999\% of the total incident power, as highlighted in Figure\ref{fig2}d. Remarkably, the disk represents only 0.44\% of the theoretical thickness limit for absorbers (refer to section 5 in Supporting Information). When subjected to $-k$ illumination, the electric field—represented in Figure \ref{fig2}d—demonstrates a slight increase in intensity but remains mostly concentrated within the cylinder's low-loss area. As a result, the backward absorption $A_{-k}$ is approximately $\sim2$\%. Despite the relatively constant magnetic field distribution within the cylinder, its intensity decreases by a factor of ten relative to the $+k$ incidence. Consequently, the exceptionally thin, asymmetric bianisotropic metasurface achieves perfect absorption. Combined with an extraordinary wavelength-to-thickness ratio of 1.13$\times 10^5$, this underscores the impressive effectiveness of the proposed structure in realizing perfect absorption with an exceptionally thin profile. 

Clearly, asymmetric reflection and absorption responses must arise from distinct total scattering cross-section (SCS) and total absorption cross-section (ACS) spectra under opposite illuminations. As shown in the top panel of Figure \ref{fig2}e, under $+k$ incidence the cylinder exhibits strong and equal SCS (red curve) and ACS (green curve) values. At the frequency of 0.602 THz, the two cross-section curves intersect at a value of 0.123 mm$^2$ -- approximately the unit cell area.
This intersection signifies a critical coupling point, enabling optimal absorption \cite{Alaee2017jpdap}. The far-field scattering pattern, as shown in the inset of the top panel of Figure \ref{fig2}e, exhibits a significant forward scattering at the frequency indicated by the dashed line, but with zero backward scattering due to the null reflection (Figure \ref{fig2}f). However, when the metasurface is illuminated in the $-k$ direction at 0.602 THz, the absorption cross-section is nearly zero (dashed green curve) while the SCS (dashed red curve) is large and near a maximum value, as shown in the lower panel of Figure \ref{fig2}e. The far-field radiation pattern under the $-k$ illumination shows similar forward and backward scattering values, as shown in Figure \ref{fig2}g, which is drastically different from that under forward incidence. Such a radiation pattern is very close to that without the conductive disk. Also noteworthy is the nearly identical forward scattering in the normal direction for both illumination directions, attributable to the structure's reciprocity.

\begin{figure}[tb]
	\centering
	\includegraphics[width=16cm]{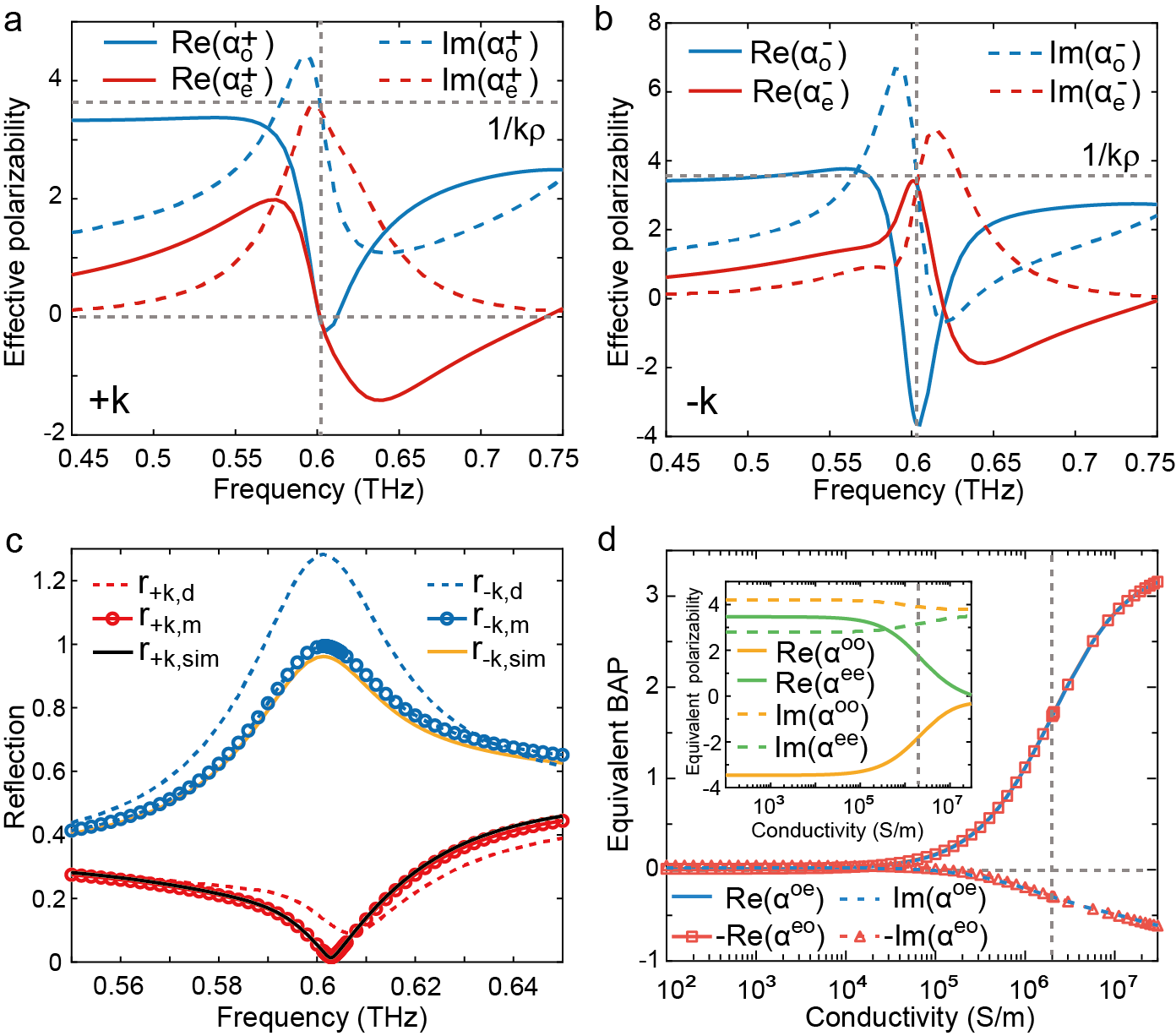}
	\caption{Multipole expansion analysis of the asymmetric MS. (a) and (b) are the calculated effective directional polarizabilities for $+k$ and $-k$ incidences, respectively. (c) The comparison of the retrieved forward and backward reflections among the simulation, the dipole model (d), and the multipole model (m). (d) Calculated equivalent multipole BAPs varying with the conductivity of the disk. The inset shows the equivalent polarizabilities as a function of the disk conductivity. }
	\label{fig3}
\end{figure}

In Figure \ref{fig1}e, we have shown that the asymmetric reflection arises from the mixing of scattering response with even and odd parity. To further elucidate the bianisotropy-induced asymmetric perfect absorption \cite{Yazdi2015tap, Hopkins2015, Odit2016apl}, we developed a bianisotropic model, which includes the multipole contributions using multipole decomposition and $\alpha$-Tensor methods \cite{Terekhov2017, Alaee2018, evlyukhin2019prb, Mun2020acsphoton}. The multipole moments are calculated using an expansion in spherical coordinates with plane waves incident in the $+k$ and $-k$ directions, respectively. It is widely recognized that the parity of multipoles, determined by spherical harmonics, can be classified as even or odd. Importantly, the parity is closely linked to the Onsager-Casimir relations \cite{Graham1997, Tretyakov2002, Kruk2016aplphoton, Dezert2019oe}. To show the cross-coupling between the even and odd multipoles, with an assumption of an $x$-polarized incident field E$_{x}$, we define two effective multipole moments based on the formulas for the normally scattered field of an arbitrarily shaped particle as \cite{Kuznetsovscirep2022, terekhov2019prb}
\begin{align}
	M_o^\pm & = p_x^\pm+(k_0/2ic) Q_{yz}^{m,\pm}  \\ 
	M_e^\pm & = m_y^\pm+(k_0c/6i) Q_{xz}^{e,\pm}  
	\label{eq:effpolar}
\end{align}
\noindent 
where $M_o^\pm(M_e^\pm)$ represents the effective moments contributed from multipoles with odd (even) parity, the superscripts + (-) indicate the incident wave propagating along $+k$ ($-k$) direction, respectively; $p_x$, $Q_{zx}^{e}$ are the induced ED and magnetic quadrupole with odd parity, respectively;  $m_y$,  and $Q_{zy}^{m}$ are the induced MD and electric quadrupole (MQ) with even parity, respectively \cite{Jackson1998}. Then the direction-dependent effective polarizabilities can be defined as $\alpha_o^\pm=M_o^\pm/\varepsilon_0E_{x}$ and $\alpha_e^\pm=\pm \eta_0M_e^\pm/E_{x}$, where $E_{x}$ is the incident electric field at the center of the mass of the cylinder, $\varepsilon_0$ is the vacuum permittivity. In the absence of a conductive disk, the cylinder exhibits two identical opposite reflections at 0.6 THz, with an amplitude of unity and a phase of -$\pi/2$ with respect to the incidence. The mid-plane of z=0 is set as the phase reference plane. By using the transmission and reflection equations based on multipole moments (Section 1 in Supporting Information), we can derive the same effective directional polarizabilities, as shown in the second row of Table \ref{tab:gb}. However, for the cylinder without the IS, the two effective moments are mixed together \cite{Poleva2023prb}. Then, under the condition of forward perfect absorption, the effective directional polarizabilities are the same with $\alpha_o+ = \alpha_e^+=\frac{1}{k_0\rho}$, where $k_0$ is the wavenumber of the resonant frequency, and $\rho=1/S_{\rm L}$ is the inverse area of the unit cell. For the condition of near-unity backward reflection, we can obtain $\alpha^-_{o}=-(\alpha^-_e)^*=\frac{-1}{k_0\rho}+\frac{i}{k_0\rho}$ as shown in the third row of the Table.\ref{tab:gb}. These conditions achieve excellent agreement with simulations, as shown in \textbf{Figure \ref{fig3}a} and b, indicated by the intersection of the dashed grey lines. We note that these conditions are also the same as those required to achieve perfect absorption and perfect reflection in the dipole limit \cite{Fan2021oe}.


The scattering behavior of subwavelength elements is primarily governed by dipoles. However, in our structure, we find that dipole contributions alone are insufficient to account for the observed phenomenon of perfect absorption and nearly perfect reflection at resonance. Rather, when the effect of multipoles is incorporated we achieve excellent agreement between retrieved values and simulation data, as displayed in Figure \ref{fig3}c (See Section 3 in Supporting Information). Further, bianisotropic perfect absorption calculated using only dipoles violates the Onsager-Casimir conditions (see Section 3 in Supporting Information). To fully describe the bianisotropic effect, the contribution of high-order multipoles cannot be easily ignored \cite{alexei2022prb, Poleva2023prb}. Incorporating $\bm{\alpha}$-tensor method \cite{Mishchenko1996, Proust2016acsphoton, Mun2020acsphoton}, we can rewrite the constitutive equations using the effective multipole moments as
\begin{equation}  \label{eq:mp} 
	\begin{split}
		\frac{1}{\varepsilon_0}M_o = \alpha^{oo}E_{x} + \alpha^{oe}(\eta_0H_{y}) \\ 
		\eta_0M_e = \alpha^{eo}E_{x} + \alpha^{ee}(\eta_0H_{y}) 
	\end{split}
\end{equation}
where $E_x$ and $H_y$ are the incident field at the position of the multipoles, and the equivalent polarizabilities under the x-polarized incidence are given as in Section 1 in Supporting Information. Based on Lorentz reciprocity \cite{Achouri2021prb} we can prove that $\alpha^{eo}=-\alpha^{oe}$, which obeys the Onsager-Casimir conditions.

The bianisotropic response of the asymmetric metasurface (MS) evidently does not solely arise due to dipole cross-coupling, and indeed the dipole to quadrupole coupling also significantly contributes to this response. If the dipoles are dominant, Equation \ref{eq:mp} can be relaxed to the standard constitutive equations with typical Onsager-Casimir conditions. Using  Eqs. (13)-(16) of Supporting Information and the Onsager-Casimir condition ($\alpha^{eo}=-\alpha^{oe}$), we obtain  {\it generalized} expressions:
\begin{align}
	t^\pm  & =1+ \frac{i k_0\rho}{2}\Big(\alpha^{oo} + \alpha^{ee} \Big) \label{eq:net} \\ 
	r^\pm  & = \frac{i k_0\rho}{2}\Big(\alpha^{oo} \pm 2\alpha^{oe}  -\alpha^{ee} \Big)\:. \label{eq:ner}
\end{align}  \noindent
These two expressions are very similar to those described using only dipoles \cite{Alaee2015prb}, however, they include the scalar polarizabilities of the effective multipole moments (\ref{eq:mp}) with odd and even parity, providing contributions to the transmission coefficient $t$ and reflection $r$ coefficient (Eqs. (13) and (14) of Supporting Information). Figure \ref{fig3}d shows the calculated equivalent polarizabilities as a function of the disk conductivity at 0.6 THz. For small values of the conductivity we find that the IS of the structure is still approximately held. The nearly perfect reflection and zero transmission enable $\rm{Re}\{\alpha^{oo}\} = -\rm{Re}\{\alpha^{ee}\} \approx -1/k_0\rho$ and $\rm{Im}\{\alpha^{oo}\} \approx \rm{Im}\{\alpha^{ee}\} \approx 1/k_0\rho$, which are consistent with those listed in the second row of Table.\ref{tab:gb}. The equivalent BAPs $\alpha^{oe}$ and $\alpha^{eo}$ are approximately zeros, which are expected. However, as the disk conductivity is larger than $10^5$ S/m, the IS is significantly broken. As a result, the bianisotropic terms become significant. Additionally, we find that the conditions $\rm{Re}\{\alpha^{oo}\} = -\rm{Re}\{\alpha^{ee}\}$ and $\rm{Im}\{\alpha^{oo}\} + \rm{Im}\{\alpha^{ee}\} = \frac{2}{k_0\rho}$, which arise from the reciprocity and near-zero transmission at the resonance, are always satisfied. As the conductivity reaches 2$\times 10^6$ S/m, i.e., forward perfect absorption state, $\rm{Re}\{\alpha^{oo}\} = -\rm{Re}\{\alpha^{ee}\} \approx \frac{-1}{2k_0\rho}$, $\rm{Im}\{\alpha^{oo}\} = \rm{Im}\{\alpha^{ee}\} \approx \frac{1}{k_0\rho}$, and $\alpha^{oe} = -\alpha^{eo} \approx \frac{1}{2k_0\rho}$, which are also consistent with the theoretical prediction as shown in Table.\ref{tab:gb}. We note that the imaginary part of the equivalent BAPs should be zero if the backward incidence is perfectly reflected, as listed in Table \ref{tab:gb}. The small deviation of the retrieved imaginary parts of the equivalent BAPs from zeros (Figure \ref{fig3}d), is caused by the imperfect backward reflection. Nonetheless, the equivalent BAPs always satisfy the Onsager-Casimir relations. 

Thus, from the general Eqs. (\ref{eq:net}), (\ref{eq:ner}),  and the above discussion, we arrive at the following scenario. The bianisotropic portions of the polarizability ($\alpha^{eo}=-\alpha^{oe}$) do not contribute to the transmission coefficient, so transmission suppression does not depend on the direction of incidence. In this case, the suppression of transmission can only be associated with the excitation of resonant modes in particles and an increase of electromagnetic fields in them. On the contrary, the reflection coefficient explicitly depends on the bianisotropic polarizabilities. For the negative direction of incidence, this contribution leads to constructive interference in the reflection and its enhancement, and for the positive direction of incidence, to destructive interference in the reflection and its significant suppression. In the latter case, the reflection is suppressed along with the transmission, which leads to the absorption of the accumulated electromagnetic energy in the particles.
\begin{table*}[tb]
	\centering
	\begin{tabular}{p{4cm}|p{4.5cm}|p{4.5cm}|p{3cm}}
		Disk radius $r_d$ & $\alpha^\pm_o,\alpha^\pm_e$ & $\alpha^{oo},\alpha^{ee}$ & $\alpha^{oe},\alpha^{eo}$\\
		\hline
		\hline 
		$r_d$=0, $r_{+k}=r_{-k}=-i \ \ \& \ \  t=0$ & $-\textrm{Re}\{\alpha^\pm_{o}\}=\textrm{Re}\{\alpha^\pm_{e}\}=\frac{1}{k_0\rho}$  $\textrm{Im}\{\alpha^\pm_{o}\}=\textrm{Im}\{\alpha^\pm_{e}\}=\frac{1}{k_0\rho}$ & $\textrm{Re}\{\alpha^{oo}\}=-\textrm{Re}\{\alpha^{ee}\}=\frac{-1}{k_0\rho}$ $\textrm{Im}\{\alpha^{oo}\}=\textrm{Im}\{\alpha^{ee}\}=\frac{1}{k_0\rho}$ & $\alpha^{oe}=\alpha^{eo}=0$  \\
		\hline
		$r_d$=$r_{opt}$, ($r_{+k}=t=0 \ \ \&\ \  r_{-k}\approx-i $)  & $\alpha^+_{o}=\alpha^+_e=\frac{i}{k_0\rho}$ \newline  $\alpha^-_{o}=-(\alpha^-_e)^*=\frac{-1}{k_0\rho}+\frac{i}{k_0\rho}$ & $\alpha^{oo}=-(\alpha^{ee})^*\approx\frac{-1}{2k_0\rho}+\frac{i}{k_0\rho}$ & $\alpha^{oe}=-\alpha^{eo}\approx\frac{1}{2k_0\rho}$ \\
	\end{tabular}
	\caption{Conditions of the effective polarizabilities at the resonant frequency of 0.6 THz to achieve symmetric perfect reflection (second row) and asymmetric perfect absorption (third row). }
	\label{tab:gb}
\end{table*}

\begin{figure}[tb]
	\centering
	\includegraphics[width=\linewidth]{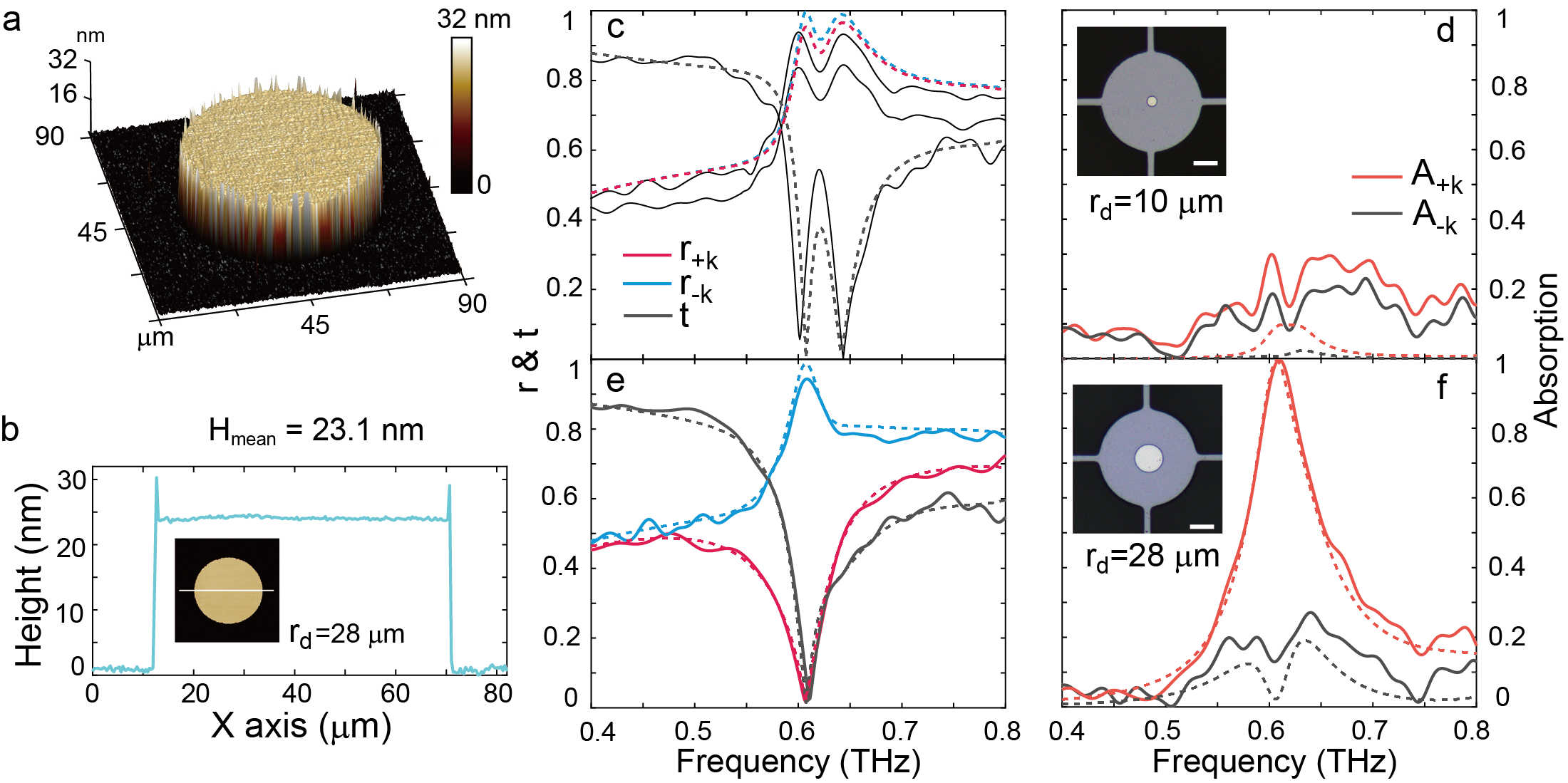}
	\caption{Experimental Results.(a) 3D view of the measured NbN disk profile using atomic force microscopy. (b) The measured height profile of the disk along the white line as indicated in the inset.  (c) \& (e) The measured (solid) and simulated (dashed) forward reflection (red), backward reflection (blue), and transmission (gray), for the samples with a disk radius of 10 and 28 $\mu$m, respectively. (d) \& (f) are the calculated forward $A_{+k}$ (red) and backward $A_{-k}$ (black) absorption spectra for the two samples. The insets are the microscopic images of the fabricated samples.}
	\label{fig4}
\end{figure}

To experimentally validate the extremely thin perfect absorber, we fabricated an array of free-standing cylinders (103 $\mu$m radius, 80 $\mu$m height, 310 $\mu$m periodicity). The high-resistive silicon cylinders are connected with 10 $\mu$m beams. Instead of gold, we utilized a 20-nm thick niobium nitride (NbN) with a room temperature conductivity of $5.5\times10^5$ S/m as the disk. (Section 6 in Supporting Information) The thickness of the NbN thin film was characterized by atomic force microscopy. \textbf{Figure \ref{fig4}a} shows the 3D view of an NbN disk after reactive ion etching. The average surface roughness of the disk is about 0.26 nm. As shown in Figure \ref{fig4}b, the average height of the disk is about 23.1 nm, which is close to the designed 20 nm. Terahertz time-domain spectroscopic measurements were performed for samples with various radii (Section 7 in Supporting Information), and Figure \ref{fig4}c-f shows the measured reflection and transmission spectra for the samples with a disk radius of 10 and 28 $\mu$m, respectively. For the $r_d$ = 10 $\mu$m sample, all scattering parameters exhibit two resonances at 0.605 THz and 0.69 THz (Figure \ref{fig4}c), which are primarily due to the weakly perturbed EH$_{111}$ and HE$_{111}$ modes, respectively. Due to the weak symmetry breaking, both reflections $r_{+k}$ and $r_{-k}$ are very similar with measured peak amplitude larger than 0.8. The calculated absorption (Figure \ref{fig4}d) for both incidences is very small since most of the power is reflected. With increasing the disk radius, the transmission $t$ at 0.605 THz does not change significantly. For $r_d$ = 28.5 $\mu$m (Figure \ref{fig4}e), the measurements exhibit a single resonance in the backward reflection $r_{-k}$ (solid blue curve) with an amplitude above 0.9 and near-zero transmission (solid black curve) at the resonance. However, the forward reflection $r_{+k}$ (solid red curve) drops to zero due to the IS breaking, leading to an asymmetric perfect absorption A$_{+k}$ at 0.605 THz (Figure \ref{fig4}f). The measured results are in excellent agreement with the simulations (dashed curves). Notably, our ultrathin perfect absorber achieves an extraordinary wavelength-to-thickness ratio exceeding 25,000, representing only about 2\% of the theoretical limit dictated by causality (Section 5 in Supporting Information).  

\begin{figure}[tb]
	\centering
	\includegraphics[width=12cm]{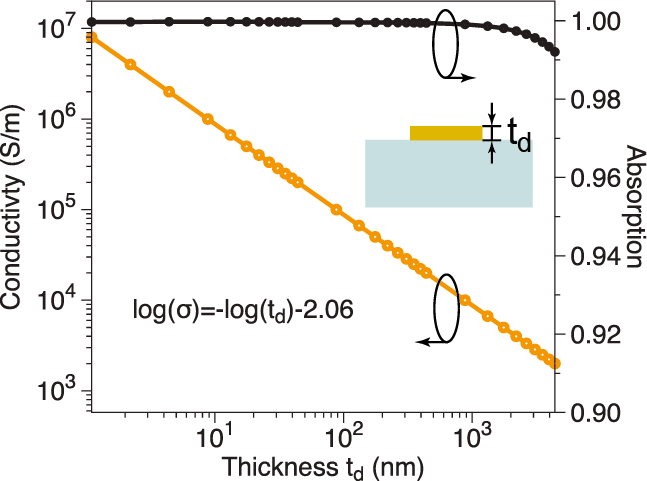}
	\caption{The relationship between the disk thickness and the metal conductivity to achieve perfect absorption at the resonance of 0.6 THz. }
	\label{fig5}
\end{figure}

Our proposed absorber demonstrates adaptability and is extendable beyond thin films and circular geometries. In \textbf{Figure \ref{fig5}} we show that it is possible to achieve perfect absorption at 0.6 THz, so long as a specific relationship between the disk conductivity $\sigma$ its thickness $t_d$ (red circles) is followed. Therefore, the surface impedance $Z_s=1/\sigma t_d$ plays an important role in attaining perfect absorption. The constancy of $Z_s$ for perfect absorption indicates a constant surface carrier concentration, assuming the carrier mobility does not change. Notably, our design with conductivity on the order of $10^7$ S/m enables the realization of atomic-level thickness for perfect absorption. Compared to other thin-film absorbers as shown in Figure \ref{fig1}(b) \cite{songadam2014,lukprb2014, Jariwalanl2016, Toudertoe2018, Horng2020prappl, Nematpouranm2021, Xiaoanm2021, Alfieriadom2022,heleraom2021}, our NbN disk exhibits the smallest VFF with a value of 1.06$\times10^{-6}$, surpassing graphene-based absorber by two orders of magnitude.

\section{Conclusion}
We have experimentally demonstrated an asymmetric absorber with the thickness of only 4 $\times 10^{-5}$ of the operational wavelength. Through breaking of the cylinder's inversion symmetry, nearly 100\% of the absorbed power was confined to the conducting disk. Our analysis reveals that such a significant asymmetric response primarily arises from the bianisotropic effect among the dipoles and quadrupoles. Additionally, we extended the Onsager-Casimir relations to encompass quadrupoles, providing a more comprehensive understanding of the system. These findings not only advance our knowledge of subwavelength particle scattering but also open up exciting possibilities for various novel applications that demand extremely confined energy, such as energy harvesting, sensitive detection, and controlled scattering manipulation.

\section{Methods}
\textbf{Numerical Simulations:} The reflection and transmission spectra are calculated using commercial electromagnetic simulation software based on the finite-element method. In the modeling, the silicon is set as a Drude-type material with a carrier density of 1$\times 10^{10}$ cm$^{-3}$. The gold disk is modeled as a Drude metal with a DC-conductivity of 2$\times 10^{6}$ S/m and a collision frequency of 2$\pi\times2\times10^{13}$ rad/s. Periodic boundaries are used for the metasurface structure. To ensure the accuracy of the simulations, the mesh layers are set to at least 4 in thickness. The multipole moments are calculated from the fields with the integration origin set as the mass center of the cylinder. Similarly, in the modeling for the NbN-based structures, the NbN thin film was modeled with a DC-conductivity of $5,5\times 10^{5}$ S/m. The far-field radiation patterns are calculated based on the electric field distribution excited by a plane wave using periodic conditions. \\
\textbf{Fabrication:} First, an ultrathin with a thickness of 20 nm NbN film was sputtered on a cleaned SOI (silicon on insulator) substrate. Then the NbN disk was defined using CF$_4$ gas in the reactive ion etching process. After removing the etching mask, the sample was coated with AZ9260 as the etching mask for following the deep reactive ion etching (DRIE) process. Next, to achieve a free-standing structure, the backside of the sample with a thickness of around 500 µm was again etched in the DRIE system. And the 1-µm buried oxide layer was removed in buffered oxide etch (BOE) solution for about 10 mins. Finally, the free-standing silicon disk array was cleaned using acetone and isopropyl alcohol.  \\
\textbf{Characterization:} Three samples with different disk radii $r_d$ were characterized by a fiber-based THz time-domain spectroscopy. In the transmission measurements, the transmission coefficient $t(\omega)$ from the samples is measured with the air as the reference. In the reflection measurement, a 3-mm high-resistive silicon is used as a beam splitter to split the incident the reflected beams. Using this setup, the terahertz beam is normally incident on the samples, which are placed on a sample mount with an open window in the center to ensure no beam is reflected from the mount to the sample. To obtain the reflection coefficient r($\omega$), a gold mirror is used as the reference. The frequency dependent absorptivity was calculated as A($\omega$) = 1-$|t(\omega)|^2$-$|r(\omega)|^2$. The disk profile was measured using the Bruker Dimension ICON. \\



\bibliography{reference}
\bibliographystyle{apsrev4-2}

\section{Acknowledgements} \par 
This work is supported by the National Natural Science Foundation of China (62275118, 62288101), the Fundamental Research Funds for the Central Universities, and the Research fund for Jiangsu Key Laboratory of Advanced Techniques for Manipulating Electromagnetic Waves; A.B.E. acknowledges support from the Deutsche Forschungsgemeinschaft (DFG, German Research Foundation) under Germany’s Excellence Strategy within the Cluster of Excellence PhoenixD (EXC 2122, Project ID No. 390833453); The work of A.E.M. was supported by the Australian Research Council (DP200101353);
W.J.P. acknowledges support from the US Department of Energy (DOE) (DESC0014372)





\end{document}